\documentclass[twocolumn,english,prl,superscriptaddress,showpacs,longbibliography,fixfloat]{revtex4-1}
\usepackage[utf8]{inputenc}
\usepackage{amsmath}
\usepackage{amssymb}
\usepackage{graphicx}

\makeatletter

\pdfpageheight\paperheight
\pdfpagewidth\paperwidth

\usepackage{xcolor}
\definecolor{darkblue}{rgb}{0.1,0.2,0.6} 
\definecolor{lightblue}{rgb}{0.1,0.1,1.0}
\definecolor{darkred}{rgb}{0.8,0.1,0.2}
\usepackage[colorlinks,citecolor=lightblue,linkcolor=darkblue,urlcolor=lightblue] {hyperref}
\renewcommand{\BibitemShut}[1]{}

\makeatother

\usepackage{babel}
\begin{document}
\global\long\def\E{\mathrm{e}}%
\global\long\def\D{\mathrm{d}}%
\global\long\def\I{\mathrm{i}}%
\global\long\def\mat#1{\mathsf{#1}}%
\global\long\def\vec#1{\mathsf{#1}}%
\global\long\def\cf{\textit{cf.}}%
\global\long\def\ie{\textit{i.e.}}%
\global\long\def\eg{\textit{e.g.}}%
\global\long\def\vs{\textit{vs.}}%
 
\global\long\def\ket#1{\left|#1\right\rangle }%

\global\long\def\etal{\textit{et al.}}%
\global\long\def\tr{\text{Tr}\,}%
 
\global\long\def\im{\text{Im}\,}%
 
\global\long\def\re{\text{Re}\,}%
 
\global\long\def\bra#1{\left\langle #1\right|}%
 
\global\long\def\braket#1#2{\left.\left\langle #1\right|#2\right\rangle }%
\global\long\def\obracket#1#2#3{\left\langle #1\right|#2\left|#3\right\rangle }%
 
\global\long\def\proj#1#2{\left.\left.\left|#1\right\rangle \right\langle #2\right|}%

\title{Multi-set matrix product state calculations reveal mobile Franck-Condon excitations under strong Holstein-type coupling}

\author{Benedikt Kloss}
\email{bk2576@columbia.edu}
\affiliation{Department of Chemistry, Columbia University, 3000 Broadway, New York,
New York 10027, USA}

\author{David Reichman}
\affiliation{Department of Chemistry, Columbia University, 3000 Broadway, New York,
New York 10027, USA}

\author{Roel Tempelaar}
\email{r.tempelaar@gmail.com}
\affiliation{Department of Chemistry, Columbia University, 3000 Broadway, New York,
New York 10027, USA}

\begin{abstract}
We show that the dynamics of (vertical) Franck-Condon excitations in the regime where Holstein-coupled vibrational modes mix strongly with electronic degrees of freedom sharply contrasts with the known self-localized behavior of vibrationally relaxed excitations. Instead, the strongly-coupled modes are found to periodically induce resonances between interacting electronic sites, during which effective excitation transfer occurs, allowing Franck-Condon excitations to attain substantial mean square displacements under conditions where relaxed excitations are essentially trapped to a single site. In demonstrating this behavior, we employ a multi-set matrix product state formalism. We find this numerically exact technique to be a remarkably efficient approach to the notoriously difficult problem posed by the Holstein model in the regime where the electronic coupling, the vibrational quantum, and the vibrational reorganization energy are comparable in magnitude.
\end{abstract}

\maketitle

\emph{Introduction}.---Holstein-type vibronic coupling, the coupling of local vibrations to the transition energies of electrons, holes, and excitons, is at the core of myriad important dynamical phenomena in the physical sciences. In addition to its importance in many inorganic systems, of particular current interest is its manifestation in organic molecular materials \cite{Spano_10a}, with implications for photosynthetic energy transfer \cite{Womick_11a, Fuller_14a, Nalbach_15a} and singlet exciton fission \cite{Bakulin_15a, Tempelaar_17a, Fujihashi_17a, Morrison_17a, Tempelaar_18a}, among many other applications. At the same time, the Holstein model continues to pose a considerable challenge for theory, encompassing a rich parameter space involving energy scales that include the electronic coupling ($J$), the vibrational energy ($\omega_0$), the vibrational reorganization energy ($g^2\omega_0$), and the temperature. While certain limits of this space are amenable to perturbative approaches \cite{Migdal_58a, Holstein_59a, Holstein_59b, Lang_62a, Feinberg_90a}, no (semi)analytical treatment is available in the regime where $J$, $\omega_0$, and $g^2\omega_0$ are comparable. This regime, where strong mixing between vibrational and electronic coordinates occurs, is representative of many organic materials, and has been the target of various numerically exact techniques that have emerged over the recent years. Such techniques are based on either an elimination of vibrational coordinates (by means of a system-bath decomposition) \cite{Tanimura_89a, Makri_95a, Diosi_97a, Chen_15a} or an explicit but truncated representation of the entire vibronic system \cite{Philpott_71a, Hoffmann_02a}. However, both approaches rapidly become prohibitively expensive with increasing number of electronic and vibrational degrees of freedom. This scaling issue drastically worsens with increasing vibrational reorganization energy, as system-bath approaches become difficult to converge and explicit descriptions demand the inclusion of an ever increasing set of bosonic states representing the vibrational coordinates. As a result of this lack of viable methodologies, much remains to be learned about how charges and excitons dynamically interact with strongly coupled vibrations.

In this letter, we employ
the remarkable computational benefits offered by tensor network states to explore the nonequilibrium excitation dynamics resulting from the single-mode Holstein model covering the full range of the vibronic coupling strength, $g$, and including the strong-mixing regime. We pay special attention to different local initial excitation conditions. For initial excitations that are vibrationally relaxed in the (shifted) excited state vibrational potential, we find the mobility to decline with increasing $g$, as expected for Holstein polarons. However, for Franck-Condon (sudden) excitations, we find the $g$-dependence to be markedly weakened for a surprisingly long time period after initialization. Concomitantly, we find the quasi-ballistic transport found in the weak coupling limit to be replaced by a pulsed transfer mechanism. An analysis of transient vibrational overlap factors shows that this mechanism is driven by a vibrational oscillation of the Franck-Condon excitation, which protects the excited state from self-localizing while allowing periodic resonances during which effective excitation transfer occurs. This mechanism allows the excitation to attain substantial mean square displacements in coupling regimes where vibrationally-relaxed excitations are essentially immobile.

Tensor network states, in particular in their matrix product state (MPS) form, have gradually attained popularity as an efficient, numerically exact framework for describing large interacting quantum systems \cite{White92a,SCHOLLWOCK201196}. Ground states of gapped one-dimensional systems are known to be efficiently representable by MPSs \cite{fannes1992,1742-5468-2007-08-P08024}. Similarly, MPSs have gained considerable traction in the application to nonequilibrium dynamics, although considerable challenges exist due to the exponential scaling of their computational cost with time for general, ergodic systems \cite{Chiara06a}. Furthermore, while ample applications can be found in strongly-correlated many-body physics, MPSs have remained relatively underrepresented in single-particle electronic problems and in particular those concerning the single-mode Holstein model, even though examples targeting the ground state of its Hamiltonian have appeared as early as two decades ago \cite{Jeckelmann_98a}. The last year has seen the appearance of a few notable works showing promising results for MPS based calculations of the Holstein model
\cite{Ren_18a, Kurashige_18a}, yet the utility of tensor network states to this class of problems has remained largely unexplored. In this letter, by demonstrating that MPSs allow access to unprecedented time and length scales for the Holstein model in the strong-mixing regime, we showcase their potential for studying a host of polaronic phenomena.

\emph{Theory}.---The Holstein Hamiltonian can be expressed in terms of $J$, $g$, and $\omega_0$ as
\begin{equation}\label{eq:Holstein}
\hat{H}= \omega_0 \sum_{\alpha=1}^N b_{\alpha}^{\dagger} b_{\alpha}  +g \omega_0 \sum_{\alpha=1}^N (b_{\alpha}^{\dagger} + b_{\alpha}) \ket{\alpha}\!\bra{\alpha} +J \!\!\!\sum_{<\alpha,\beta>}^N\!\!\! \ket{\alpha}\!\bra{\beta},\nonumber
\end{equation}
where $b_\alpha^{(\dagger)}$ is the annihilation (creation) operator for a local mode coupled to an electronic excitation $\ket{\alpha}$ at site $\alpha$,
and the last summation is limited to nearest-neighboring sites, $\alpha$ and $\beta$.
Note that this Hamiltonian includes local coupling of each electronic site to a single, dispersionless Einstein oscillator. More general coupling schemes would pose no difficulty for the applied methodology, but are beyond the scope of the present study.

Tensor network states employ the principle that the wavefunction coefficients of a state in a Hilbert space for $N$ particles can be thought of as a tensor of order $N$. Decomposing this tensor into a product of smaller tensors, and truncating these tensor products, replaces the exponential scaling with $N$ by a low polynomial (usually linear) scaling.
In case of MPSs, such a decomposition takes the form
\begin{equation}\label{eq:MPS_simple}
\ket{\Psi}=\sum_{\{\sigma_i\}}A_{1}^{\sigma_{1}}A_{2}^{\sigma_{2}}\dots A_{N}^{\sigma_{N}} \ket{ \sigma_{1}\sigma_{2} \dots \sigma_{N}},
\end{equation}
where the indices $\sigma_i$ label the physical basis states, and the matrices satisfy $A_{i}^{\sigma_{i}}\in\mathbb{C}^{\chi_{i-1}\times\chi_{i}}$. Here, the ``bond dimension'' $\chi_i$ controls the truncation applied to the internal (virtual) indices. The degree of entanglement between bipartitions of the system that can be accounted for by a tensor network is determined by its bond dimensions as well as its connectivity. Nonequilibrium dynamics generally leads to an exponential increase of the bond dimension necessary to describe the state accurately with time, with the exception of localized systems \cite{Bardarson2012a}. Thus, it is crucial to select a tensor network ansatz that captures the entanglement build-up efficiently in order to simulate physically relevant time scales.

For the Holstein model, an obvious choice for the tensor network ansatz is to consider the Hamiltonian as a chain of spinless noninteracting fermions, each of which is coupled to its respective vibrational mode. After performing a Jordan-Wigner transformation on the fermions, this problem can straightforwardly be treated as an MPS. However, the reachable timescale under this ansatz is limited due to the relatively fast growth of entanglement entropy. An alternative approach considers the Holstein model as an $N$-level impurity, the levels of which correspond to the electronic single-particle states, where each level is coupled to its respective mode. Within the setting of MPSs, this impurity model is treated as an effectively one-dimensional problem. However, the resulting connectivity introduces long-range interactions between the vibrational coordinates and the impurity site, again leading to a rapid growth of entanglement.


Our tensor network ansatz is closely related to the $N$-level impurity approach, but instead of solving the entire system as a single MPS, we expand the total wavefunction in terms of a set of products of electronic states and associated vibrational wavefunctions,
\begin{equation}\label{eq:MSMPS_all}
\ket{\Psi}=\sum_{\alpha=1}^{N}\ket{\Psi^{\alpha}}\ket{\alpha}.\nonumber
\end{equation}
The vibrational wavefunctions are then each expanded independently as an MPS, analogous to Eq.~\ref{eq:MPS_simple}, the norm of which corresponds to the electronic population at the associated site. The index ${\sigma_i}$ then labels the vibrational states using the bosonic occupation number basis associated with the (unshifted) ground state harmonic potential, which are truncated beyond a maximum number of quanta, $\nu_\text{max}$. In this choice of ansatz, the Hamiltonian contains only uncorrelated terms for the physical degrees of freedom, such that entanglement is introduced only indirectly and at a slower rate than in the other possible choices mentioned above. Despite introducing a quadratic scaling in the system size, we find it to achieve remarkably long length and time scales. Note that such a \emph{multi-set} approach was first introduced \cite{Fan_94a,Wor96a} for multi-configuration time-dependent Hartree methods \cite{MEYER199073,BECK20001}, a related tensor network technique, and was very recently employed in an MPS setting close in spirit to the one applied here \cite{Kurashige_18a}.

To obtain the time evolution of $\ket{\Psi}$, we use the time-dependent variational principle \cite{dirac_1930,frenkel1934mechanics}, which allows the computation of a (time-local) optimal approximate solution to the time-dependent Schr\"odinger equation, given a variational ansatz (such as the multi-set MPS as employed in this work) \cite{Haegeman11a,Haegeman_16a}. It amounts to solving the projected Schr\"odinger equation,
\begin{equation}\label{eq: TDVP}
i\dot{\ket{\Psi}}=\mathcal{P_M}[\ket{\Psi}] \hat{H} \ket{\Psi},\nonumber
\end{equation}
where $\mathcal{P_M}[\ket{\Psi}]$ is the projector onto the tangent space of the variational manifold $\mathcal{M}$ attached to  $\ket{\Psi} \in \mathcal{M}$ \cite{Haegeman13a}. The dynamics is numerically exact up to times for which the variational ansatz fails to capture the produced entanglement accurately, and can be systematically converged to longer times by increasing the bond dimension. All presented data is tightly ($<1\%$ deviation) converged with respect to boundary effects as well as all numerical parameters, which include the bosonic truncation $\nu_\text{max}$ (up to 128) and the bond dimension $\chi_i$ (32 in all cases). Applying less stringent convergence criteria is tempting; however, there is numerical evidence that loose convergence of asymptotic properties can yield qualitatively incorrect results \cite{Kloss18a}.

Not being limited to ground state or band-edge excited states, we are free to differentiate between the following two vibrational initial conditions. The first condition is that of an excitation relaxed in the (electronically) excited state vibrational potential (referred to as ``relaxed''), whereas the second corresponds to an excitation created upon a vertical transition starting from the zero-phonon (electronic) ground state level (known as a Franck-Condon excitation). These two cases can be regarded as the two extremes spanning the scope of commonly used nonequilibrium initial conditions. The Franck-Condon excitation is representative of an impulsive optical excitation of a vibronic system involving a vibration whose energy is large compared to the thermal quantum, which is satisfied by most functionally-relevant Holstein modes studied in the literature. The relaxed excitation, on the other hand, is a pragmatic initial condition for models involving a ``shifted'' basis for describing (electronically) excited state vibrations \cite{Hestand_15a}, and can be regarded as a proxy for optical pumping into the lowest-energy vibronic ($0-0$) transition.

\begin{figure}
\includegraphics{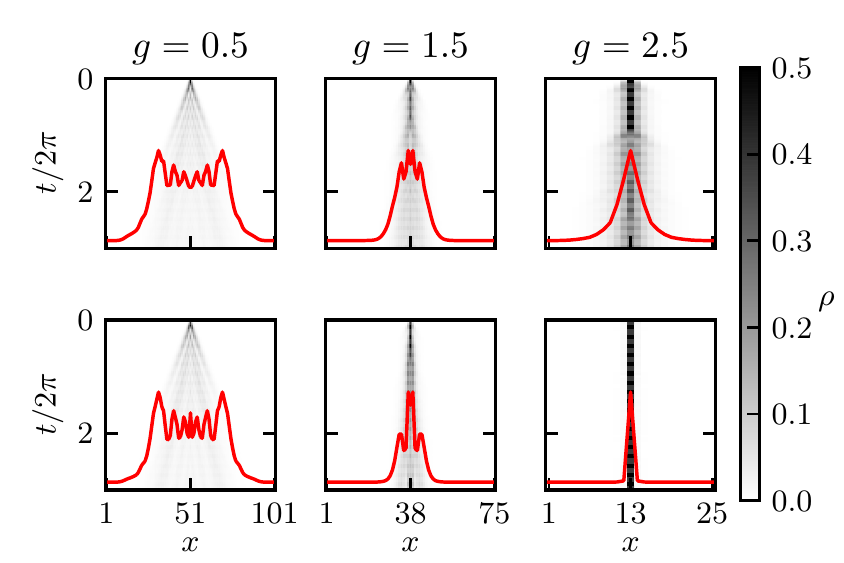}
\caption{\label{fig:profiles} Excitation density  $\rho$ as a function of time (vertical) and site (horizontal) for Franck-Condon (upper panels) and relaxed (lower panels) excitations.
Red curves show the excitation density profile at $t/2\pi=3$.}
\end{figure}

\begin{figure}
\centering
\includegraphics{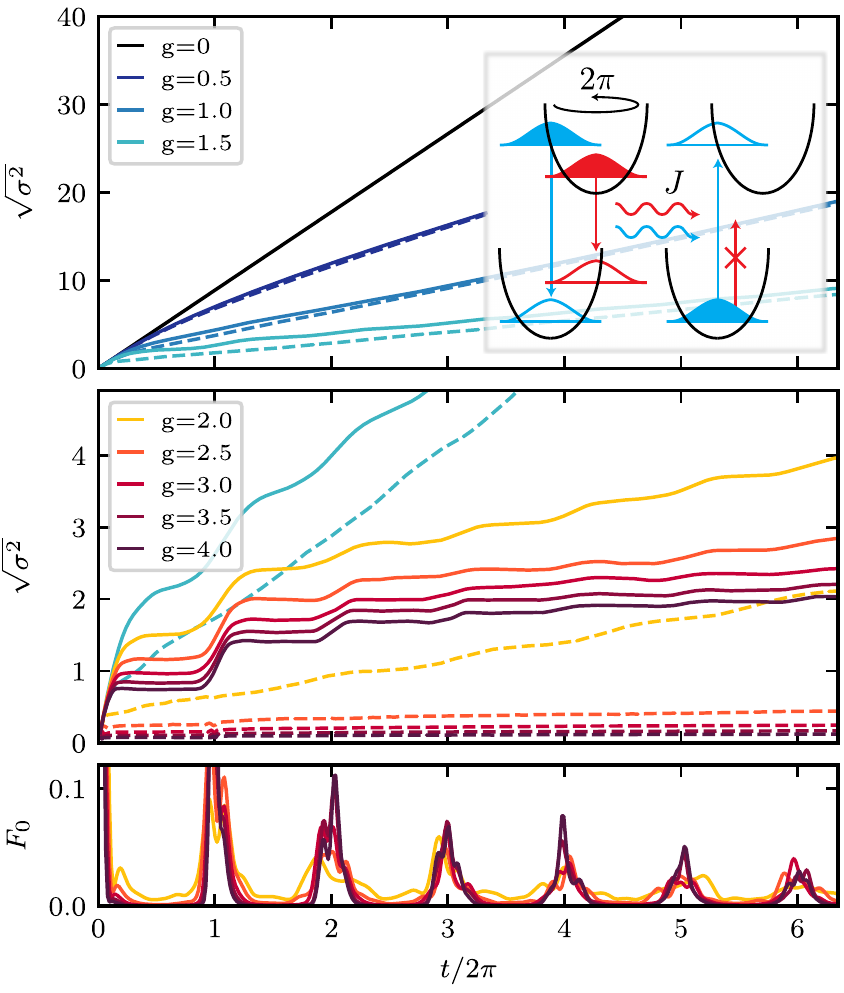}
\caption{\label{fig:MSD} Upper panels: Root mean square displacement against time for Franck-Condon (solid lines) and relaxed (dashed lines) excitations. The inset shows a schematic of the vibrationally induced transfer of Franck-Condon excitations. Data for $g=1.5$ reproduced in both panels for comparison. Lower panel: Overlap $F_0$ between the vibrational wavepacket in the electronically excited potential and that of the zero phonon state in the ground state potential, averaged over electronic sites.}
\end{figure}

\emph{Results}.---In Fig.~\ref{fig:profiles} we present excitation dynamics for a linear chain with open boundaries, following an initial excitation located at the chain center, under the two aforementioned vibrational preparations. Shown as heatmaps are the calculated chain populations as a function of time (in units of inverse energy, with $\hbar=1$) resulting from the Holstein model with $\omega_0=J=1$ and for varying values of $g$. The dynamics for $g=0.5$ is near-identical for both initial vibrational conditions, which is consistent with the notion that these conditions become equivalent in the limit of $g\rightarrow 0$, and is dominated by a ballistic component characteristic of a vibronically uncoupled excitation. The excitation mobility can be seen to decrease with increasing $g$, indicative of the formation of a polaron with increasing effective mass, including a rapid decline of the mobility of a relaxed excitation in the regime of strong coupling, as a result of self-localization. However, in marked contrast to the relaxed excitation, the Franck-Condon excitation is seen to retain a substantial mobility even under strong coupling. This trend is shown more systematically in Fig.~\ref{fig:MSD}, which depicts the transient root mean square displacements for values of $g$ ranging from 0 to 4. For $g=2.5$, the Franck-Condon excitation spread rapidly reaches $\sim$6 sites, whereas the relaxed excitation remains essentially stalled on a single site.


\begin{figure}
\centering
\includegraphics{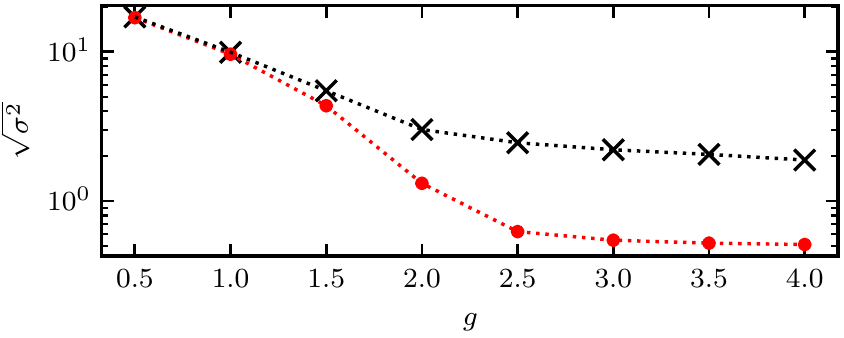}
\caption{\label{fig:MSD_trend} Root mean square displacement at time $t/2\pi=3$ for Franck-Condon (black crosses) and relaxed (red dots) excitations.}
\end{figure}

An alternative means of demonstrating the contrasting dynamics emerging from relaxed and Franck-Condon excitations is by plotting the root mean square displacement at a fixed time for varying $g$. This is shown in Fig.~\ref{fig:MSD_trend}, for $t/2\pi=3$. Here, the delocalization of the relaxed excitation shows a pronounced drop with $g$ exceeding unity, which is almost entirely absent for the Franck-Condon excitation. Interestingly, within this coupling range, we see the emergence of a beating pattern for the Franck-Condon excitation dynamics in Fig.~\ref{fig:MSD}, with $2\pi$-periodic enhancements in the root mean square displacement becoming more abrupt with increasing $g$. This indicates that their dynamics stems from a mechanism that is radically different from that of relaxed excitations in the strong coupling limit. In order to understand the nature of this mechanism, it is insightful to consider the transient overlap of the vibrational wavefunction inside the electronically excited potential with that of the zero phonon state in the ground state potential. Shown together with the root mean square displacements in Fig.~\ref{fig:MSD}, this overlap exhibits a beating pattern roughly in sync with that seen for excitation transport, such that regions of large vibrational overlap coincide with abrupt enhancements in the mean square displacement.

The physical picture of the dynamics of strongly-coupled Franck-Condon excitations emerging from our results is shown schematically in the inset of Fig.~\ref{fig:MSD}. Upon initial excitation, the vibrational wavepacket oscillates in and out of the Franck-Condon region, as indicated by the beatings apparent in the calculated vibrational overlap. When inside this region, excitation transfer to neighboring sites is effective due to a resonance between the (inverted) Franck-Condon transition at the donor site and that at the neighboring site. Moving out of this region, however, the transition energy at the donor site will rapidly decrease, leaving the vibrationally relaxed neighboring without an energy-matching transition with significant vibrational overlap. Moreover, the sustained motion of the strongly-coupled vibration protects the electronic excitation from self-localizing while such periodic resonant transfers occur.

\begin{figure}
\includegraphics{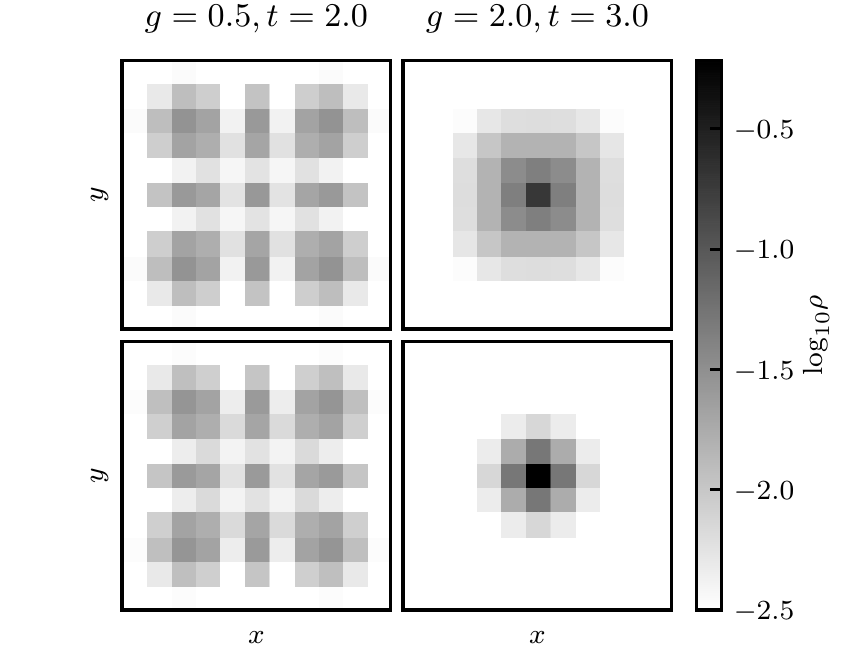}
\caption{\label{fig:profiles2D} Excitation density $\rho$ for a two-dimensional lattice. Upper (lower) panels show Franck-Condon (relaxed) excations. Log scale is used for clarity. 
}
\end{figure}


Next, we briefly discuss the manifestation of the dynamics in two dimensions. Higher-dimensional tensor network structures are computationally demanding, while mapping higher-dimensional problems to a one-dimensional tensor network induces long-range connectivity, resulting in complicated entanglement structures. This renders the application of tensor network states in two dimensions and above a challenging endeavor, particularly so for dynamical problems involving strongly-correlated electronic systems for which the time scales reached in state-of-the-art calculations have been limited \cite{Zaletel15a}. Interestingly, for the two-dimensional Holstein model, again with $\omega_0=J=1$, we are able to reach comparable time scales ($t\sim3J)$ with modest computational resources.
Although a detailed investigation of the performance of our method in higher dimensions is beyond the scope of the present letter, we speculate that the principle that entanglement is introduced indirectly (by coupling between different sets) aids in obtaining a favorable computational performance even in two dimensions.
In Fig.~\ref{fig:profiles2D}, we show the excitation densities for selected reorganization energies and excitation conditions, given a square lattice with an excitation initiated at the center. For weak coupling, both the relaxed and Franck-Condon excitation are similarly spread-out, exhibiting a well-resolved two-dimensional interference pattern. Consistent with the one-dimensional case, with increasing $g$ we find the spread of the Franck-Condon excitation to be much more pronounced than that of the relaxed excitation.

\emph{Conclusions}.---We have shown that mixing of electronic coordinates with strongly coupled vibrational modes results in Franck-Condon excitations whose initial dynamics is markedly different from that known for vibrationally relaxed excitations. Sustained vibrational motion is found to generate periodic resonances between neighboring electronic sites, during which effective energy transfer occurs, allowing a Franck-Condon excitation to spread over substantial distances in parameter regimes where relaxed excitations are essentially self-trapped on a single site. Of course, over much longer time scales one expects that this mechanism no longer governs the dynamics, and behavior akin to that of the relaxed initial condition takes over. In addition to providing fundamental insights into strongly interacting vibronic systems, these results have implications for the nonequilibrium behavior of materials upon vertical transitions from a vibrationally relaxed (ground state) initial condition, in particular when the functional material dimensions are in the range of the mean square displacements found in our calculations. In many practical cases excitation conditions are near-vertical, resulting from impulsive perturbations of the electronic degrees of freedom, and our findings reveal that the subsequent ultra-fast dynamics can not be understood based on vibrationally-relaxed steady-state principles. In unraveling this remarkable photophysical behavior, we have addressed a notoriously difficult region of the Holstein model, employing the computational benefits offered by a multi-set matrix product state approach. As such, this work highlights the potential of this approach in addressing problems involving charged and excitonic polarons. The flexibility of the approach also allows to make progress in more complex situations, such as long-range electronic hopping or higher dimensionality, for which encouraging results have been presented in this work.

\emph{Acknowledgements}.---The authors thank Miles Stoudenmire for helpful discussions and Joonho Lee for pointing to a recent work employing a similar method \cite{Kurashige_18a}. D.R.R. and B.K. acknowledge funding from NSF Grant No. CHE-1839464.
\bibliographystyle{apsrev4-1}
\bibliography{Bibliography}

\end{document}